# Mapping and manipulation of topological singularities: from photonic graphene to T-graphene

Sihong Lei[1†], Shiqi Xia [1†], Junqian Wang[1], Xiuying Liu[2], Liqin Tang[1], Daohong Song[1, 3*], Jingjun Xu[1], Hrvoje Buljan[1, 4*], and Zhigang Chen[1, 3*]

*1 The MOE Key Laboratory of Weak-Light Nonlinear Photonics, TEDA Applied Physics Institute and School of Physics, Nankai University, Tianjin 300457, China*

*2 College of Physics and Materials Science, Tianjin Normal University, Tianjin 300387, China*

*3 Collaborative Innovation Center of Extreme Optics, Shanxi University, Taiyuan, Shanxi 030006, China*

*4 Department of Physics, Faculty of Science, University of Zagreb, Bijenička c. 32, Zagreb 10000, Croatia*

†*These authors contributed equally to this work*

*e-mail: zgchen@nankai.edu.cn, hbuljan@phy.hr, songdaohong@nankai.edu.cn

**Abstract:** Topological singularities (TSs) in momentum space give rise to intriguing fundamental phenomena as well as unusual material properties, attracting a great deal of interest in the past decade. Recently, we have demonstrated universal momentum-to-real-space mapping of TSs and pseudospin angular momentum conversion using photonic honeycomb (graphene-like) and Lieb lattices. Such mapping arises from the Berry phase encircling the Dirac or Dirac-like cones, and is thus of topological origin. In this paper, we briefly present previous observations of topological charge conversion, and then we present our first theoretical analysis and experimental demonstration of TS mapping in a new T-graphene lattice. Unlike other lattices, there are two coexisting but distinct TSs located at different high-symmetry points in the first Brillouin zone of T-graphene, which enables controlled topological charge conversion in the same lattice. We show active manipulation of the TS mapping, turning the two TSs into vortices of different helicities, or one into a high-order vortex but the other into a quadrupole. Such TS manipulation and pseudospin-to-orbital conversion may find applications in optical communications and quantum information, and may bring insight into the study of other Dirac-like structures with multiple TSs beyond the 2D photonic platform.

Topological phases of matter, typically described by the topological band theory using concepts such as the Chern number, Berry phases and polarization, have become a rapidly growing field in condensed matter physics and interdisciplinary studies, following the seminal work by Kane and Mele regarding topological insulators or quantum spin Hall insulators[1]. Indeed, topological phases and pertinent properties including topologically protected boundary states have been demonstrated in a variety of electronic, photonic, acoustic, mechanical and ultracold atom systems[2-8].

Optical vortices have drawn continued and ever-increasing attention in recent years, especially with advances in nanophotonic technologies for realization of phase and polarization singularities[9-11]. This is largely due to numerous applications of optical angular momentum (OAM) arising from optical vortices, including optical manipulation[12, 13], optical communications[14, 15] and quantum information processing[16]. In most previous studies, optical vortices are typically generated with spatial light modulators, q-plates metasurfaces, or active vortex generators[17-19]. On the other hand, a Dirac point in the momentum space as that displayed by graphene can be considered as a topological singularity (TS) with nontrivial Berry phase winding. In recent experiments, it has been demonstrated that optical vortices can been generated by mapping the TSs from momentum to real space using photonic structures[20-27]. This brings about a new way for generation, manipulation, and transmission of optical vortex beams; along with new features mediated by pseudospin-to-orbital angular momentum conversion[20, 21, 23, 25, 28].

Artificial honeycomb lattices (HCLs) have served as an ideal model for exploring fundamental phenomena in graphene and Dirac-like materials[29]. In photonics, waveguide arrays are arranged to form HCLs, namely, the photonic graphene[30], where light propagation can mimic electronic transport properties in graphene. Based on such photonic graphene lattices, we have observed several intriguing phenomena including unconventional edge states[31], valley Landau-Zener-Bloch oscillations[32], valley-dependent vortex states[24] and wavepacket self-rotation[27], in addition to the pseudospin angular momentum[21]. In particular, under proper alignment of the pseudospin states near the Dirac cones, optical vortices of different topological charges can be generated via momentum-to-real-space TS mapping[25]. The basic scheme is illustrated in Figure 1, where Figure 1a is for the half-integer pseudospin ($S = 1/2$) such as with the HCL, whereas Figure 1b is for the integer pseudospin ($S = 1$) such as with the Lieb lattice. Unlike the spin in electronic systems[1, 33], the pseudospin angular momentum is not associated with any intrinsic property of particles, but rather comes from spatial sublattice degrees of freedom[21, 25, 28]. Therefore, different pseudospin modes (e.g., $s = \pm 1/2$ for HCL and $s = \pm 1, 0$ for Lieb lattice) can be selectively excited, leading to controlled topological charge conversion.

In this work, we present the first theoretical analysis and experimental demonstration of TS mapping in a new kind of photonic lattices, namely, the T-graphene lattice. It was theoretically proposed for the study of transport properties of electrons in 2D materials[34-37], since such a lattice exhibits features of Dirac-like fermions similar to graphene but without the HCL structure[38]. In addition, the appearance of nearly flat-band states with nonzero Chern number in T-graphene has drawn much attention, and it is expected to be a good candidate for single-layer elemental superconductors[39]. Unlike previously studied HCL or Lieb lattices, there are two distinct (well separated in the eigenvalue spectrum) TSs coexisting at different high-symmetry points (HSPs) of the T-graphene lattice. We show active manipulation of the TS mapping by selective excitation of the HSPs, which leads to high-order vortex generation of different helicities, or turns the component of pseudospin mode into quadrupole when the excited TS is selectively eliminated.

First, let us briefly review the topological charge conversion as a way for vortex generation in photonic graphene (Figure 2a1) and Lieb lattices (Figure 2c1) via a single TS mapping (Figure 2a2 and Figure 2c2). Mathematically, the wave dynamics around a Dirac point in such Dirac-like structures is governed by the effective Hamiltonian[25, 28]

$$\mathcal{H}(\boldsymbol{p}) = \mu I + \kappa(S_x p_x + S_y p_y) \qquad (1)$$

where $\mu, \kappa$ depend on the potential and couplings of the lattice, $I$ is an identity matrix, and $p_x$ and $p_y$ are the displacements with respect to the TS, $S_i$ are the components of the pseudospin angular momentum operator $\boldsymbol{S}$. The Hamiltonian $\mathcal{H}$ has $2S + 1$ degenerated bands at the TS point. For instance, the photonic graphene and Lieb lattice host a doubly and a triply degenerated TS, respectively, which offer an ideal platform to demonstrate topological charge conversion between different pseudospin modes experimentally. An optimal topological charge conversion occurs with the rule $l \to l + 2s$, where $l$ and $s$ are, respectively, the topological charge of the vortex beam and the optimally aligned pseudospin state[25], as illustrated in Figure 2a2 and Figure 2c2.

Our previous experimental results are summarized in Figure 2, where the photonic graphene and Lieb lattices fabricated using multi-beam optical induction technique are shown in Figure 2b1, and

Figure 2d1)[21, 25]. A vortex beam with a topological charge $l=1$ (see inset in Figure 2b2) is sent into the photonic graphene to excite the pseudospin mode $s=1/2$, and the output interferogram clearly displays a topological charge $l=2$ (Figure 2b2). In the Lieb lattice, on the other hand, the probe beam excitation ($l=1, s=1$) leads to the generation of output vortex with a topological charge $l=3$ (Figure 2d2). These output results exhibit a conversion rule of the topological charge going from $l$ to $l+2s$ for the optimally aligned excitations. Theoretically, the Berry phase winding ($w\pi$) around the TS is found to be responsible for the topological charge conversion ($w=1$ for the photonic graphene and $w=2$ for the Lieb lattice), which is robust even in systems where angular momentum is not fulfilled. The optimal topological conversion obeys $l \to l+w$ for $l>0$ and $l \to l-w$ for $l<0$[25]. This conversion rule contains the topological quantity $w$, and is found to be more general than the one containing pseudospin $l \to l+2s$, indicating the topological origin and universality for the TS mapping[25].

Thus far, the study of TS mapping has been limited to the single TS excitation in HCL ($S=1/2$) or Lieb ($S=1$) lattices (Figure 2)[20-22, 25], but it has not been demonstrated for coexisting higher order TSs located at different HSPs as those manifested by the T-graphene lattice (Figure 3b). In real space, T-graphene is complex square lattice with four sites in one unit-cell (Figure 3a). When we consider the tight-binding model without the next-nearest-neighbor (NNN) couplings, the T-graphene has two coexistent TSs located at two different HSPs: $\Gamma$ and $M$ (Figure 3b). As we shall show from theoretical analysis and experimental demonstration, controllable topological charge conversion can be realized under judicious excitation of different TSs in a T-graphene lattice. Moreover, two TSs can be selectively eliminated by introducing the NNN couplings in the lattice (Figure 3c, d), which offers a new platform for studying the interaction and manipulation of TSs.

To analyze the underlying physical mechanism, we consider the tight-binding Hamiltonian of T-graphene lattice

$$H(\mathbf{k}) = \begin{pmatrix} 0 & t_1 & t_2\exp(ik_xa_1)+t_3 & t_1 \\ t_1 & 0 & t_1 & t_2\exp(-ik_ya_2)+t_3 \\ t_2\exp(-ik_xa_1)+t_3 & t_1 & 0 & t_1 \\ t_1 & t_2\exp(ik_ya_2)+t_3 & t_1 & 0 \end{pmatrix} \quad (2)$$

where $t_1, t_2, t_3$ are intra-cell, inter-cell and NNN coupling, respectively. We set the magnitudes of the basic vectors $\mathbf{a_1}$ and $\mathbf{a_2}$ (see Figure 3a) to be 1 so to simplify the following discussion. In order to have an explicit solution of pseudospin eigenmodes, the "redundant" Bloch band for a triply degenerated TS at the HSPs of $H(\mathbf{k})$ should be removed, such as the upper band that shall not be considered for the TS at $\Gamma$ point shown in Figure 3b. We first expand $H(\mathbf{k})$ around the HSPs. Under Taylor expansion, elements in eq 2 can be expressed as $\exp(i(k_i + p_i)) = \mathcal{A}(1 + ip_i + o(p_i))$, where $p_i$ is the displacement of transverse wavevector with respect to HSPs, and $i = x, y$, $\mathcal{A} = \exp(ik_i)$. When $k_i$ is located at $\Gamma$ point with $\mathbf{k}^\Gamma = (0,0)$, $\mathcal{A}$ is equal to 1, induced by the momentum shift between the HSPs. Likewise, when $k_i$ is located at $M$ point with $\mathbf{k}^M = (\pi, \pi)$, $\mathcal{A}$ is equal to $-1$.

Moreover, we can choose the basis $\mathcal{S} = (\psi_1, \psi_2, \psi_3, \psi_4)$, where $\psi_1 = (i, -1, -i, 1)^T/2$, $\psi_2 = (1, -1, 1, -1)^T/2$, $\psi_3 = (1,1,1,1)^T/2$, $\psi_4 = (i, 1, -i, -1)^T/2$ will be identified as the pseudospin eigenmodes in the following discussion, to perform a unitary transform of $H_{HSP}(\mathbf{p})$, which yields

$$\mathcal{H}(\mathbf{p}) = \mathcal{S}^\dagger H_{HSP}(\mathbf{p})\mathcal{S} = \frac{1}{2}\begin{pmatrix} -2t_2\mathcal{A} - 2t_3 & t_2\mathcal{A}pe^{-i\theta} & t_2\mathcal{A}pe^{i\theta} & 0 \\ t_2\mathcal{A}pe^{i\theta} & -4t_1 + 2t_2\mathcal{A} + 2t_3 & 0 & t_2\mathcal{A}pe^{-i\theta} \\ t_2\mathcal{A}pe^{-i\theta} & 0 & 4t_1 + 2t_2\mathcal{A} + 2t_3 & t_2\mathcal{A}pe^{i\theta} \\ 0 & t_2\mathcal{A}pe^{i\theta} & t_2\mathcal{A}pe^{-i\theta} & -2t_2\mathcal{A} - 2t_3 \end{pmatrix} \quad (3)$$

where $p_x + ip_y = pe^{i\theta}$. At $\Gamma$ point, $\mathcal{A} = 1$ and $\text{diag}(\mathcal{H}) = (-t_2 - t_3, -2t_1 + t_2 + t_3, 2t_1 + t_2 + t_3, -t_2 - t_3)$. The diagonal terms denote the eigenvalues of $\psi_i$, which indicates the degeneracy of Bloch eigenmodes as well as the possible existence of TS. When the NNN coupling $t_3$ is set to zero, $t_1 = \pm t_2$ guarantees degeneracy of eigenmodes. Because the couplings of photonic T-graphene in our experiments are positive, we set $t_1 = t_2 = t > 0$, and there will be three degenerated modes $\psi_1, \psi_2,$ and $\psi_4$ at $\Gamma$ point. Meanwhile, the large difference between diagonal terms in eq. 3 suppresses the energy exchange between $\psi_3$ and the three degenerated modes. Therefore, elements related to $\psi_3$ can be neglected and the dynamics of excitation around Dirac-like cones at $\Gamma$ point is governed by the effective Hamiltonian $\mathcal{H}_\Gamma(\mathbf{p}) = -tI + \frac{1}{\sqrt{2}}tp_x S_x^\Gamma + \frac{1}{\sqrt{2}}tp_y S_y^\Gamma$, where

$$S_x^\Gamma = \frac{1}{\sqrt{2}}\begin{pmatrix} 0 & 1 & 0 \\ 1 & 0 & 1 \\ 0 & 1 & 0 \end{pmatrix}, S_y^\Gamma = \frac{1}{\sqrt{2}}\begin{pmatrix} 0 & -i & 0 \\ i & 0 & -i \\ 0 & i & 0 \end{pmatrix}, S_z^\Gamma = \begin{pmatrix} 1 & 0 & 0 \\ 0 & 0 & 0 \\ 0 & 0 & -1 \end{pmatrix} \quad (4)$$

are pseudospin angular momentum operators and $I$ is the identity matrix. The eigenmodes of pseudospin $\chi_s^\Gamma$ are given by $S_z^\Gamma \chi_s^\Gamma = s\chi_s^\Gamma$, which are $\chi_+^\Gamma = \psi_1$, $\chi_0^\Gamma = \psi_2$, $\chi_-^\Gamma = \psi_4$. On the other hand, at $M$ point, $\text{diag}(\mathcal{H}) = (t_2 - t_3, -2t_1 - t_2 + t_3, 2t_1 - t_2 + t_3, t_2 - t_3)$., which has same TS degeneracy condition $t_1 = \pm t_2$ when the NNN couplings $t_3$ is set to zero. Elements related to $\psi_2$ will be removed. We also set $t_1 = t_2 = t > 0$, $\mathcal{H}_M$ around $M$ point is reduced to $\mathcal{H}_M = tI + \frac{1}{\sqrt{2}} tp_x S_x^M + \frac{1}{\sqrt{2}} tp_y S_y^M$, which has the similar expression as $\mathcal{H}_\Gamma$, where $S_x^M = S_x^\Gamma$, $S_y^M = -S_y^\Gamma$. This leads to different conversion of pseudospin eigenmodes $\chi_-^M = \chi_+^\Gamma$, $\chi_+^M = \chi_-^\Gamma$ at different TSs in the T-graphene ($\chi_-^M$ denotes the pseudospin eigenmodes at $M$ point with eigenvalue $s = -1$).

When the intra- and inter-cell couplings are identical $t_1 = t_2$, with negligible NNN coupling $t_3 = 0$, this symmetry dictates simultaneous presence of two TSs. However, by proper adjustment of the couplings that breaks this symmetry, one can tune the existence of one or the other TS. When $t_2 = t_1 + t_3, t_3 \neq 0$, the TS is preserved at $M$ point but eliminated at $\Gamma$ point (Figure 3c). The other relation $t_2 = t_1 - t_3, t_3 \neq 0$ reverses the TS degeneracy condition (Figure 3d), where the TS at $\Gamma$ point is retained but that at $M$ point is not. The selective elimination of one TS offers a new scheme for regulating topological charge conversion from the momentum to real space.

In order to show the TS mapping at different HSPs and the distinction between preserved and eliminated TSs explicitly, we analyze the overlap between eigenmodes of Bloch band and pseudospin eigenmodes. Intriguingly, the Bloch eigenmodes at the TS have a fixed overlap with a given pseudospin eigenmodes, which can serve as a fingerprint of the presence of a TS[40]. We define the mode overlap as

$$\gamma_{\pm,n}(\boldsymbol{k}) = |\langle \chi_\pm | \phi_n(\boldsymbol{k}) \rangle|^2 \qquad (5)$$

where $\phi_n(\boldsymbol{k})$ is the eigenmode of $n$-th Bloch band at a given Bloch vector. The overlaps $\gamma_{\pm,n}(\boldsymbol{k})$ and eigenvalues $\beta$ of the corresponding Bloch modes are plotted in Figure 4 ($n = 1, 2, 3, 4$ for the four Bloch bands of the T-graphene lattice, and $\chi_\pm$ is the eigenmode of pseudospin $s = \pm 1$). The mode overlaps $\gamma_{\pm,n}(\boldsymbol{k})$ can be used to reveal the evolving trend of each band between different TSs, where the band away from the TSs at HSPs has no contribution on the pseudospin dynamics (top band at $\Gamma$ and bottom band at $M$ in Figure 4a). More importantly, the calculated fingerprint values $\gamma_{\pm,n}$ for three pertinent Bloch bands $\phi_n(\boldsymbol{k})$ at TS are (0.25 0.5 0.25), which can serve as an indicator of the existence of TSs even one of them is selectively eliminated. When we have $t_2 = t_1 + t_3$ by considering the NNN couplings, TS, indicated by the fingerprint values of top three bands, remains at $M$ point (Figure 4b) but vanishes at $\Gamma$ point. The pseudospin eigenmodes are no more the superposition of Bloch modes of the three degenerated bands. On the other hand, TS can be selectively eliminated at $M$ point, when the relations of coupling are set to $t_2 = t_1 - t_3$ (Figure 4c). The overlaps $\gamma_{0,n}(\mathbf{k})$ between Bloch bands and pseudospin zeros mode $\chi_0$ under three different conditions illustrated in SM also imply the TS transition selective elimination at different HSPs[40].

More interestingly, when both TSs are present, the same pseudospin eigenmodes can result in the opposite topological charge conversion in real space. Even if one of the TSs is selectively eliminated, the other TS can still maintain the relevant topological characteristics. On the one hand, the conservation of $[J_z, H] = 0$ guarantees the topological charge conversion from momentum to real space around the TS, where $J_z$ is the z-component of the total angular momentum defined by $\mathbf{J} = \mathbf{L} + \mathbf{S}$ (where $\mathbf{L} = \mathbf{r} \times \mathbf{k}$ is the orbital angular momentum). In other words, if an initial probe beam carrying a topological charge ($l$) aims at one of the z-components of the pseudospin $s$, it will convert to $l'$ at pseudospin $s'$ after propagation through lattice, and all of the components obey $l + s = l' + s'$. Since the output components are intertwined on all sublattices, with a proper excitation, we can realize an optimal TS mapping $l \to l + 2s$ at one TS. Meanwhile, the topological charge conversions are protected by the topological phase circling around TSs. When a single pseudospin component is initially excited, since pseudospin eigenmode can be seen as the superposition of all Bloch modes related to Dirac-like cone (Figure 4), Berry phase embedded in pseudospin eigenmodes will be uncovered during the evolving of wave and eventually manifest as the topological charges in real space[25]. The opposite winding of Berry phase at different HSPs in T-graphene can be used to reverse the helicity of the optical vortices after optimal topological conversion, and once converted, they maintain robust even when the other TS is eliminated.

Next, we present our experimental demonstration of different topological charge conversion under coupling condition $t_1 = t_2$. Photonic T-graphene lattices are established by multi-beam optical induction method[41, 42]. Several quasi-plane waves with proper phase and different $k$-directions are used to form writing beams for the T-graphene lattice (Figure 5a1)[40], which is invariant throughout a 20mm-long SBN crystal. A positive electric field ($180 kVm^{-1}$) applied along the c-axis of crystal translates the lattice intensity pattern into refractive index change via the photorefractive effect[43, 44]. The lattice spacing is about $22\mu m$ as shown in Figure 5a2. The probe beam with an overall Gaussian envelope is constructed by spatial light modulator (SLM) to match the phase distribution of $\psi_1$ at different HSPs (Figure 5b1, c1), and it travels through the 20mm long crystal, captured at the back facet of crystal. Since all components of the probe beam carry the same topological charge[25], the interferogram from one of the components delivers the phase information of the whole beam. When the probe beam

($l = 0$) is launched to $M$ point in momentum space, it excites the $s = -1$ pseudospin eigenmode $\chi_-^M$ and exports a topological charge $l' = -2$ on account of nontrivial winding of the Berry phases at corresponding TS point (Figure 5b2). In contrast to that, as the probe beam aims at $\Gamma$ point, the initial excited pseudospin component turns into $\chi_+^\Gamma$ even though $\psi_1$ of initial probe beams is unchanged (Figure 5c1). As a result, the probe beam evolves into a vortex beam with an opposite topological charge $l' = 2$ (Figure 5c2). Other experiment results with different initial probe beams $\psi_4$ are further discussed in the SM, where the probe beam initially excites either $s = 1$ or $s = -1$ component at $M$ or $\Gamma$ point. Therefore, owing to different pseudospin eigenmode excitation, opposite topological charge conversion occurs at different HSPs[40]. These results clearly indicate that the TS mapping and vortex conversion is achieved by controlled excitation in the T-graphene lattice.

Finally, to substantiate the topological charge conversion in T-graphene presented above, we numerically simulate beam propagation corresponding to experiments based on the paraxial wave equation:

$$i\frac{\partial}{\partial z}\Psi(x,y,z) = -\frac{1}{2k_0}\nabla_\perp^2 \Psi(x,y,z) - \frac{k_0 \Delta n(x,y)}{n_0}\Psi(x,y,z) \quad (6)$$

where $\Psi(x,y,z)$ is the complex amplitude of the linearly polarized electric field of probe beams; $k_0 = 2\pi n_0/\lambda$ is the wavenumber in the crystal; $n_0$ is the ambient refractive index of the medium; $\Delta n$ is the index change which denotes the T-graphene lattice array, and $z$ in eq 6 indicates the propagation distance mathematically equivalent to the role of time in quantum mechanics. The simulation results based on parameters close to those used in experiment are mapped to the pseudospin eigenmodes according to their mode distributions on sublattices at different TSs. The numerical simulations reveal that the topological charge appears in each pseudospin component of probe beams (Figure 5b3-b5, c3-c5), which agrees well with theoretical analysis and experiment results. From the phase structure of each component, we find that, if the $s = -1$ component at $M$ point is initially excited, the topological charge emerging in the $s' = 1$ component will be $l' = -2$ (Figure 5b3). For the latter two cases ($s' = 0$, $s' = -1$), the initial vortex is transformed into topological charge of $-1$ (Figure 5b4) or $0$ (Figure 5b4), while all components satisfy the angular momentum conservation $l + s = l' + s'$. Likewise, the input beam ($l = 0$) that initially excites the $s = 1$ component at $\Gamma$ point obeys the conversion rule (Figure 5c3-c5). The same rule can be found in $\psi_4$ with Gaussian envelope, which initially excites $s = 1$ ($s = -1$) at $M$ ($\Gamma$) point. Decomposition of the outputs again proves that the angular momentum conservation is satified[40]. The experiment results with initial probe beams $\psi_2$ and $\psi_3$, exciting $s = 0$ at $\Gamma$ and $M$ points, respectively, are further discussed in SM, which are all consistent with the results stated above[40].

As to the selective elimination of TSs, previous pseudospin mode excitation can directly map the condition of TS from momentum to real space. Since the elimination requires the NNN couplings in T-graphene, we choose pseudospin modes $\chi_-^M$ and $\chi_-^\Gamma$ with Gaussian envelopes to numerically excite $M$ and $\Gamma$ points utilizing discrete model. During propagation, pseudospin components always maintain a ring-shape as long as the initially excited TS is retained. Such ring-shape outputs with topological charge $l' = 2$ can be observed at $s' = 1$ components when $\chi_-^M$ ($\chi_-^\Gamma$) excites T-graphene systems which have a TS at $M$ ($\Gamma$) point (Figure 6a1, b1 and Figure 6a2, c2). In contrast, the pseudospin component $s' = 1$ splits into a quadrupole under $s = -1$ excitation of the HSP where the TS is eliminated (Figure 6c1, b2). Therefore, all the numerical results support that the TSs can be selectively and independently eliminated. The topological charge conversion indicates the presence of TS, whereas the quadrupole distribution of the pseudospin component at the output indicates eliminated TS at the pertinent HSPs.

In conclusion, we have theoretically analyzed and experimentally demonstrated different topological charge conversions in the T-graphene photonic lattices exhibiting two TSs at different HSPs in momentum space, and we have shown that the two TSs can be selectively eliminated by introducing the NNN coupling. We have put forward a method to exclude the redundant or irrelevant bands in analyzing the TS mapping, which should be applicable to the analysis of other systems with TSs, such as for multiple pseudospin components in super-honeycomb lattices[45] and quadratic band touching in Kagome lattices[46]. The overlapping between Bloch and pseudospin modes is used to reveal the TS transition at different HSPs, which provides fingerprint values to distinguish the TSs. We have also found that the fingerprint values, and more importantly, the associated topological characteristics persist at one TS when the other TS is selectively eliminated, which may promote related research on other complex systems where one can take the advantage of vortex generation from universal TS mapping. The concept and the methods developed here with respect to pseudospin manipulation and TS mapping may prove relevant to the studies of similar phenomena in other branches of topological physics[31, 47-50], and may be further extended to complex momentum-space topological singularities such as knots and links as those in Weyl lattices in higher-dimensional systems[51-57].

**Acknowledgments**

We acknowledge financial support from the National Key R&D Program of China (2017YFA0303800), the National Natural Science Foundation of China (12134006, 11922408), the Natural Science Foundation of Tianjin (21JCJQJC00050), and the QuantiXLie Center of Excellence, a project co-financed by the Croatian Government and the European Union through the European Regional Development Fund the Competitiveness and Cohesion Operational Programme (KK.01.1.1.01.0004). S. X. acknowledges support from the China Postdoctoral Science Foundation (BX2021134, 2021M701790).

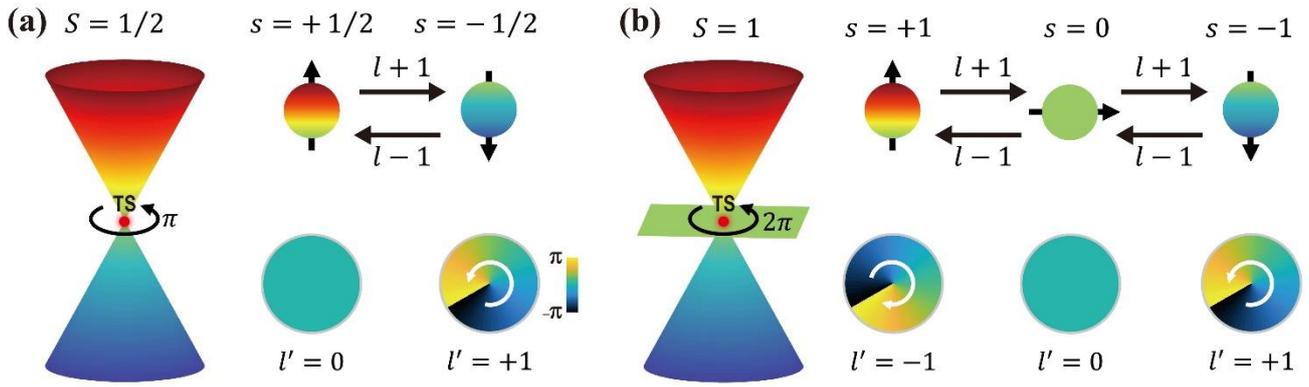

**Figure 1: Illustration of different pseudospin modes in photonic lattices.** (a) TS with half integer pseudospin $S = 1/2$, where two pseudospin components are included, i.e., $s = +1/2$ (upward arrow) and $s = -1/2$ (downward arrow). The black arrow circling around the TS (around the conical intersection) denotes the winding of the Berry phase The (vortex) topological charge for corresponding pseudospin components is shown by their phase distribution (the topological charge $l$ at $s = +1/2$ is set to 0). (b) TS with integer pseudospin $S = 1$ with associated three pseudospin components and topological charge phase distribution are sketched analogously as in (a). Here $S$ and $s$ denote the total and *z*-component of the pseudospin angular momentum, respectively. White arrows indicate the gradient of phase in vortex from $-\pi \to \pi$.

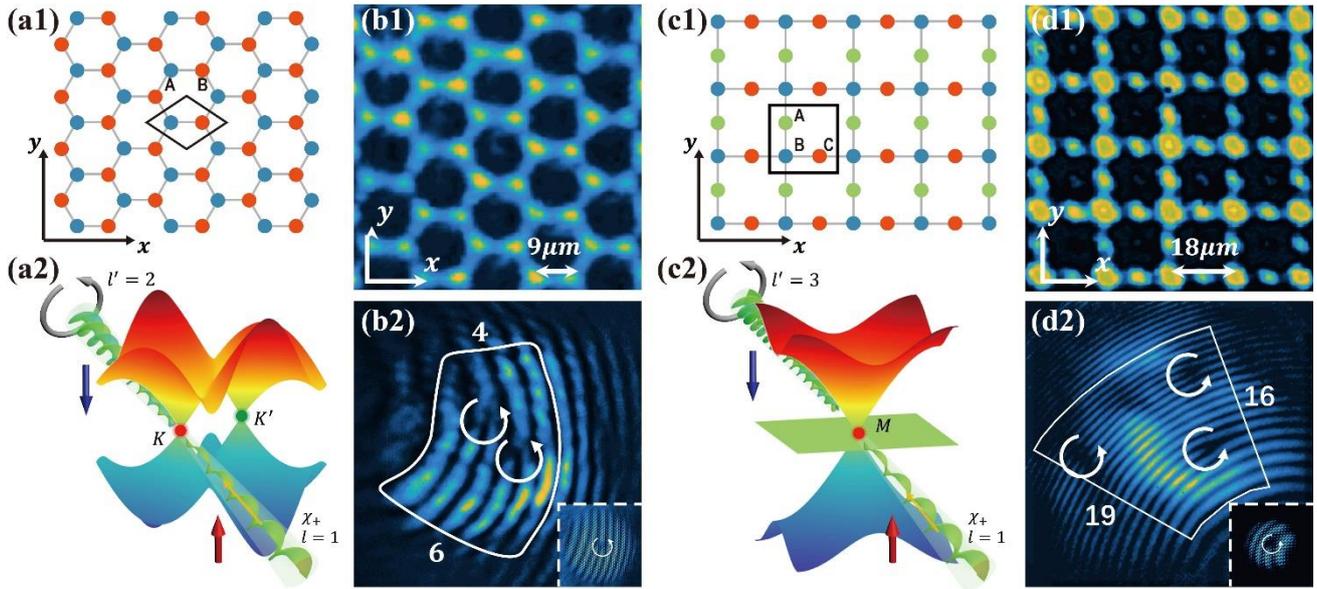

**Figure 2: Mapping of a single topological singularity in photonic graphene and Lieb lattices.** (a) Demonstration of half integer pseudospin mode excitation in photonic graphene which has two sublattices A and B (a1). Vortex beams excite modes around the conical intersections at $K$ point in the Brillouin zone (a2). (b1) Photonic graphene established in experiments by optical induction technique. Excitation of pseudospin state $s = 1/2$ with initial vortex beams $l = 1$ (inset in b2), leading to topological charge conversion from 1 to 2 (b2) as illustrated in (a2). (c, d) has the same layout with (a, b) but with integer pseudospin $s = 1$ in Lieb lattice which has three sublattices (c1). White curved arrows mark the position and helicity of the vortices.

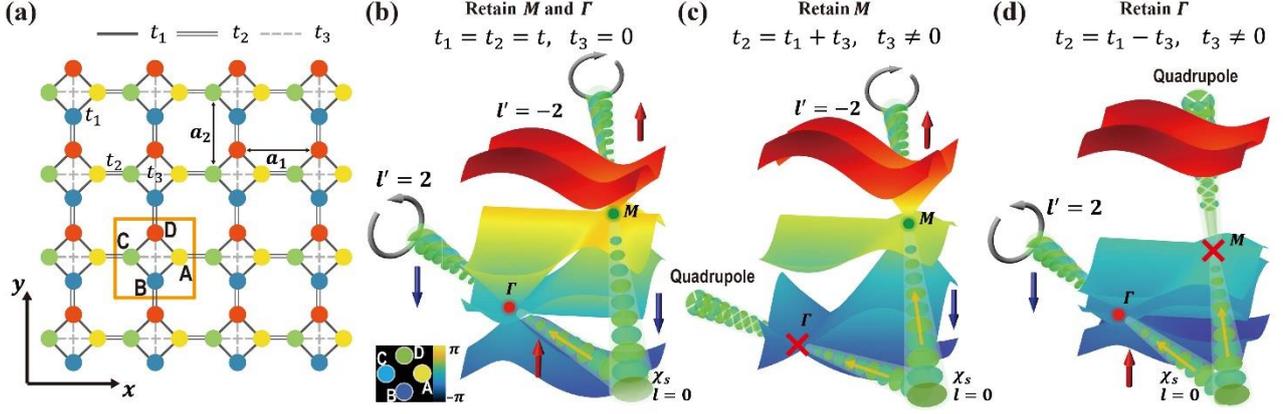

**Figure 3 Illustration of mapping and selective elimination of TSs in T-graphene lattices.** (a) The structure of a photonic T-graphene lattice in real space, where the unit-cell is illustrated by a square, and $t_1$, $t_2$ and $t_3$ are defined here as intra-cell, inter-cell and the NNN couplings, respectively. $\boldsymbol{a_1}$ and $\boldsymbol{a_2}$ are lattice vectors. (b-d) Pseudospin eigenmode $\chi_s$ with Gaussian envelope whose phase distribution is shown in inset is sent to excite $\varGamma$ and $M$ point in momentum space. (b) The band structure has two coexistent TSs located at $\varGamma$ and $M$ points in the absence of NNN coupling $t_3$. When $\chi_s$ aims at $M$ in momentum space, it excites the $s = -1$ (blue arrow) pseudospin eigenmode and acquires $l' = -2$ on $s' = 1$ (red arrow) at the output. Excitation at $\varGamma$ undergoes the opposite topological charge conversion: initial $\chi_s$ will match $s = 1$, which results in $l' = 2$ on $s' = -1$ at the output. (c, d) For parameters where one of the TS is eliminated as shown in (c) and (d), the beam will evolve into a quadrupole at $\varGamma$ point (c) or the $M$ point (d). The topological charge conversion still occurs at the other HSP where the TS is retained ($M$ point in (c) or $\varGamma$ point in (d)). We choose $t_1 = t_2 = 1$ in all panels. $t_3 = 0$ in (b) and $t_3 = 0.5$ in (c, d).

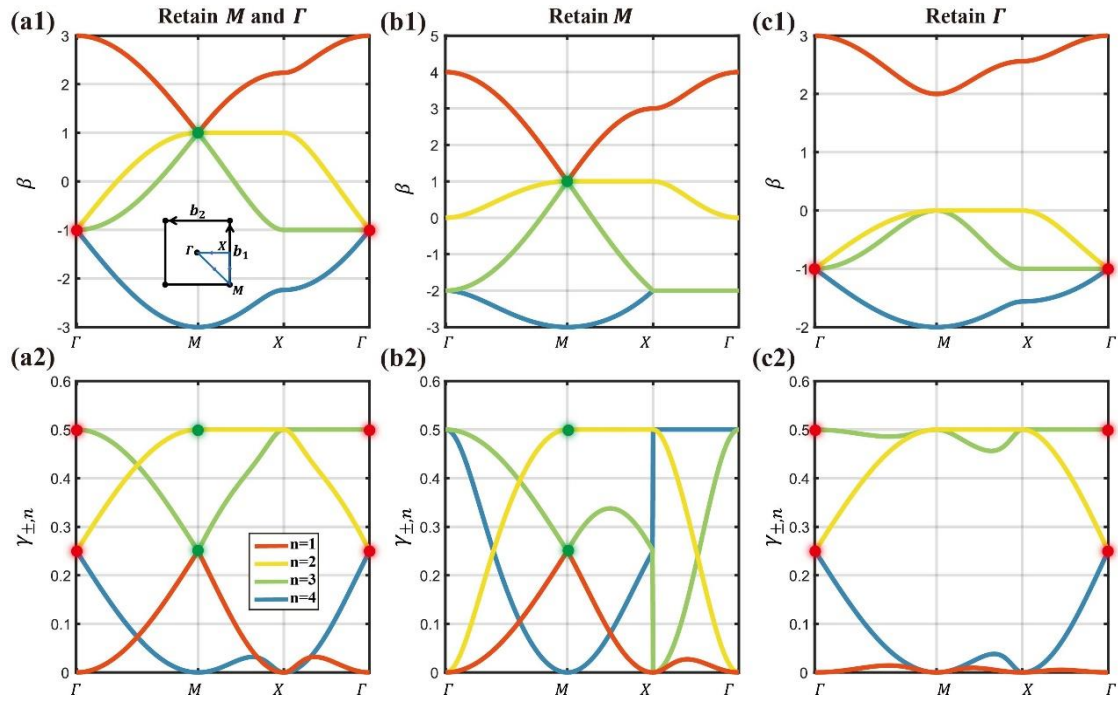

**Figure 4 Band structures and pseudospin mode occupation in T-graphene lattices of different bonding.** (a1) Band structure of T-Graphene lattice along the HSPs ($\Gamma \to M \to X \to \Gamma$) denoted by blue arrows in inset under the presence of both TSs, corresponding to the 3D band illustration in Figure 3b. $\boldsymbol{b_1}, \boldsymbol{b_2}$ are vectors of the reciprocal basis, $\beta$ are the eigenvalues of the Bloch modes. Red and green dots mark the position of TSs. (a2) The distribution of $\gamma_{\pm,n}(\boldsymbol{k})$ (mode overlap between pseudospin modes $s = \pm 1$ and Bloch modes $\phi_n(\boldsymbol{k})$) along the HSPs. (b) and (c) have the same layouts as (a), corresponding to the 3D band illustration in Figure 3c, d, respectively. The values of $t_1$, $t_2$ and $t_3$ are the same as in Figure 3.

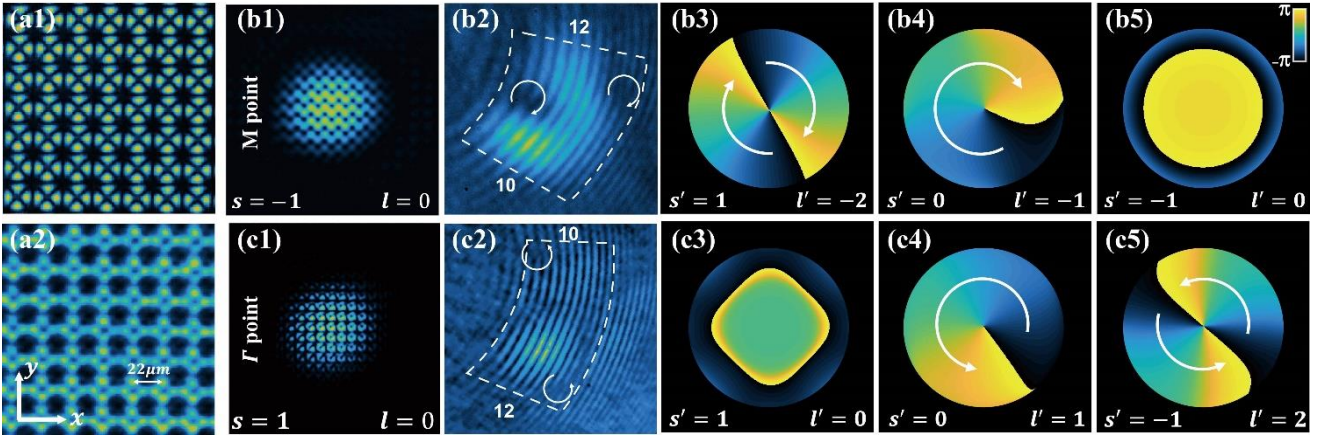

**Figure 5 Experimental demonstration of TS mapping and vortex generation in photonic T-graphene.** (a1) is the lattice beams constructed by SLM. (a2) shows an optically induced T-Graphene lattice in experiment. Pseudospin mode $\psi_1$ ($l=0$) with Gaussian envelope is sent to excite pseudospin states $s=-1$ at $M$ point (b1), $s=1$ at $\Gamma$ point (c1). Output interferograms (b2, c2) show opposite topological charge conversions. Difference in the numbers of counted fringes from the two sides of the white square region illustrates the net topological charges at output. White curved arrows mark the position and helicity of the vortices. (b3-b5, c3-c5) show output phase structure of the probe beam numerically decomposed for each pseudospin component $s'$, where corresponding topological charge $l'$ at output in each component has been identified. These results correspond to the scenario illustrated in Figure 1b.

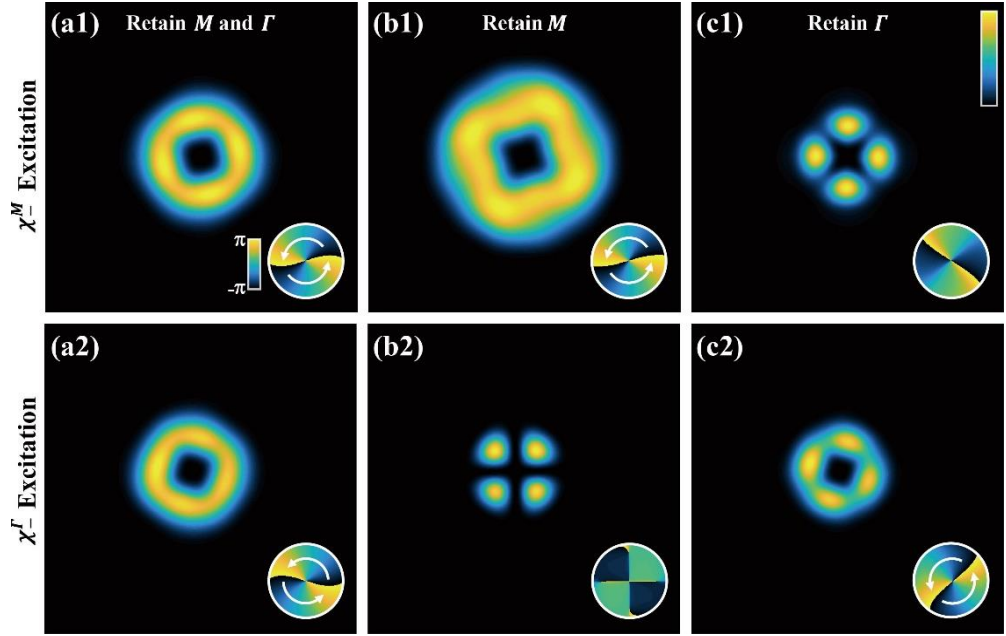

**Figure 6 Numerical calculation of output pseudospin components in T-graphene lattices of different bonding.** (a) Output results at $s' = 1$ components under $s = -1$ pseudospin mode excitation at $M$ (a1) and $\Gamma$ (a2) points, where both TSs are retained corresponding to Figure 4a. The inset shows associated phase distribution. (b) and (c) have the same layouts as (a), corresponding to Figure 4b, c, respectively. The intensities of outputs split into quadrupole in which the TS is selectively eliminated. The values of $t_1, t_2$ and $t_3$ are the same with Figure 4.